\documentclass[twocolumn,showpacs,preprintnumbers,amsmath,amssymb]{revtex4}

\usepackage{graphicx}
\usepackage{dcolumn}
\usepackage{bm}

\usepackage{subfig}
\usepackage{comment}

\begin{document}


\title{Weak Ties: Subtle Role in the Information Diffusion in Online Social Networks}

\author{Jichang Zhao}
 \affiliation{State Key Laboratory of Software Development Environment, \\
 Beihang University, Beijing 100191, P.R.China}
\author{Junjie Wu}%
 \affiliation{%
Information Systems Department, School of Economics and Management,
\\Beihang University, Beijing 100191, P.R.China
}%
\author{Ke Xu}
 \email{kexu@nlsde.buaa.edu.cn}
 \affiliation{ State Key Laboratory of Software Development Environment, \\Beihang University, Beijing 100191, P.R.China
}%

\date{\today}

\begin{abstract}
As a social media, online social networks play a vital role in the social
information diffusion. However, due to its unique complexity, the
mechanism of the diffusion in online social networks is different from the
ones in other types of networks and remains unclear to us. Meanwhile, few
works have been done to reveal the coupled dynamics of both the structure
and the diffusion of online social networks. To this end, in this paper,
we propose a model to investigate how the structure is coupled with the
diffusion in online social networks from the view of weak ties. Through
numerical experiments on large-scale online social networks, we find that
in contrast to some previous research results, selecting weak ties
preferentially to republish cannot make the information diffuse quickly,
while random selection can achieve this goal. However, when we remove the
weak ties gradually, the coverage of the information will drop sharply
even in the case of random selection. We also give a reasonable
explanation for this by extra analysis and experiments. Finally, we
conclude that weak ties play a subtle role in the information diffusion in
online social networks. On one hand, they act as bridges to connect
isolated local communities together and break through the local trapping
of the information. On the other hand, selecting them as preferential
paths to republish cannot help the information spread further in the
network. As a result, weak ties might be of use in the control of the
virus spread and the private information diffusion in real-world
applications.
\end{abstract}

\maketitle

\section{\label{sec:introduction}Introduction}

The emergence of the Internet has changed the way of communication
radically and, especially, the development of Web 2.0 applications
has led to some extremely popular online social sites, such as
Facebook~\cite{facebook}, Flickr~\cite{flickr},
YouTube~\cite{youtube}, Twitter~\cite{twitter},
LiveJournal~\cite{livejournal}, Orkut~\cite{orkut} and
Xiaonei~\cite{xiaonei}. These sites provide a powerful means of
sharing information, finding content and organizing
contacts~\cite{mislove-2007-socialnetworks} for ordinary people.
Users can consolidate their existing relationships in the real world
through publishing blogs, photos, messages and even states. They
also have a chance to communicate with strangers that they have
never met on the other end of the world. Based on the development
and prevalence of the Internet, online social sites have reformed
the structure of the traditional social network to a new complex
system, called the online social network, which attracts a lot of
research interests recently as a new social media.

Recent works about online social networks mainly focus on probing and
collecting network
topologies~\cite{mislove-2007-socialnetworks,cyworld-www07}, structural
analysis~\cite{mislove-2007-socialnetworks, cyworld-www07,
fengfu-eaofonline, fengfu-socialdelima}, user
interactions~\cite{mc-socialcascade, bv-evulutionoffacebook,
rhythmsofinteraction} and content generating
patterns~\cite{rhythmsofinteraction, guolei-kdd09}. At the same time, some
concepts and methods of traditional social networks have also been
introduced into current researches: The strength of ties is one of them.
The strength of ties was first proposed by Granovetter in his landmark
paper~\cite{tie-strength-define} in 1973, in which he thought the strength
of ties could be measured by the relative overlap of the neighborhood of
two nodes in the network. It was interesting that different from the
common sense, he found that loose acquaintances, known as weak ties, were
helpful in finding a new job~\cite{finding-job}. This novel finding has
become a hot topic of research for decades. In ~\cite{predict-model}, a
predictive model was proposed to map social media data to the tie
strength. In ~\cite{mobile-network}, Onnela et al. gave a simple but
quantified definition to the overlap of neighbors of nodes $i$ and $j$ as
follows:
\begin{equation}
\label{eq:tiestrength} w_{ij}=\frac{c_{ij}}{k_i-1+k_j-1-c_{ij}},
\end{equation}

\noindent where $c_{ij}$ is the number of common acquaintances, $k_i$ and
$k_j$ are the degrees of $i$ and $j$, respectively. {\bf In this paper, we
define $\boldsymbol{w_{ij}}$ as the strength of the tie between
$\boldsymbol{i}$ and $\boldsymbol{j}$}. The lower $w_{ij}$ is, the weaker
the strength of tie between $i$ and $j$ is.

As a social media, the core feature of online social networks is the
information diffusion. However, the mechanism of the diffusion is
different from traditional models, such as
Susceptible-Infected-Susceptible (SIS), Susceptible-Infected-Recovered
(SIR)~\cite{epidemicspreading, infectiondynamics} and random
walk~\cite{randomwalks, y-exploring, c-exploring}. At the same time, few
works have been done to reveal the coupled dynamics of both the structure
and the diffusion of online social networks~\cite{threshold-model,
measurement-flick}. To meet this critical challenge, in this paper, we aim
to investigate the role of weak ties in the information diffusion in
online social networks.

By monitoring the dynamics of
\begin{equation}
\label{eq:averages} \bar{S}=\sum_{S<S_{\max}}\frac{nS^2}{N},
\end{equation}

\noindent where $n$ is the number of connected clusters with $S$
nodes, and $N$ is the size of the network, a phase transition was
found in the mobile communication network during the removal of weak
ties first~\cite{mobile-network}. We find that this phase transition
is pervasive in online social networks, which implies that weak ties
play a special role in the structure of the network. This
interesting finding inspires us to investigate the role of weak ties
in the information diffusion. To this end, we propose a model
$ID(\alpha,\beta)$ to characterize the mechanism of the information
diffusion in online social networks and associate the strength of
ties with the process of spread. Through the simulations on
large-scale real-world data sets, we find that selecting weak ties
preferentially to republish cannot make the information diffuse
quickly, while the random selection can. Nevertheless, further
analysis and experiments show that the coverage of the information
will drop substantially during the removal of weak ties even for the
random diffusion case. So we conclude that weak ties play a subtle
role in the information diffusion in online social networks. We also
discuss their potential use for the information diffusion control
practices.

The rest of this paper is organized as follows. Section~\ref{sec:datasets}
introduces the data sets used in this paper. In
Section~\ref{sec:sroleofweakties}, we study the structural role of weak
ties. The model $ID(\alpha,\beta)$ is proposed in
Section~\ref{sec:droleofweakties}, and the role of weak ties in the
information diffusion is then investigated. Section~\ref{sec:dcontrol}
discusses the possible uses of weak ties in the control of the virus
spread and the private information diffusion. Finally, we give a brief
summary in Section~\ref{sec:summary}.

\section{Data Sets}\label{sec:datasets}

We use two data sets in this paper, i.e., \texttt{YouTube} and
\texttt{Facebook} in New Orleans. \texttt{YouTube} is a famous video
sharing site, and \texttt{Facebook} is the most popular online social site
which allows users to create friendships with other users, publish blogs,
upload photos, send messages, and update their current states on their
profile pages. All these sites have some privacy control schemes which
control the access to the shared contents. The data set of
\texttt{YouTube} includes user-to-user links crawled from \texttt{YouTube}
in 2007~\cite{mislove-2007-socialnetworks}. The data set of
\texttt{Facebook} contains a list of all the user-to-user links crawled
from the New Orleans regional network in \texttt{Facebook} during December
29th, 2008 and January 3rd, 2009~\cite{bv-evulutionoffacebook}. In both
two data sets, we treat the links as undirected.

\begin{figure}[ht]
\centering
\includegraphics[scale=0.55]{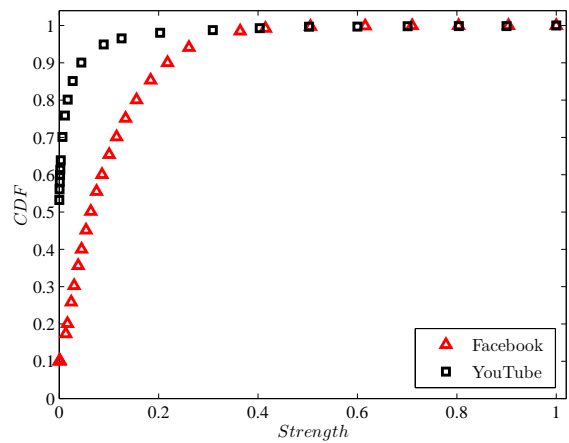}
\caption{\label{fig:cdfofstrength}(Color online) $CDF$ of the strength of
ties.}
\end{figure}

In these data sets, each node represents a user, while a tie between two
nodes means there is a friendship between two users. In general, creating
a friendship between two users always needs mutual permission. So we can
formalize each data set as an undirected graph $G (V, E)$, where $V$ is
the set of nodes and $E$ is the set of ties. We use $\arrowvert V
\arrowvert$ to denote the size of the network, and $\arrowvert E
\arrowvert$ to denote the size of ties. Some characteristics of the data
sets are shown in Table~\ref{tab:datasets}. The $Cumulative\ Distribution\
Function (CDF)$ of the strength of ties is shown in
Fig.~\ref{fig:cdfofstrength}.
\begin{table}
\caption{\label{tab:datasets}Data Sets}
\begin{ruledtabular}
\begin{tabular}{ccc}
Data set&$\arrowvert V \arrowvert$&$\arrowvert E \arrowvert$\\
\hline
\texttt{YouTube} & 1134890 & 2987624\\
\texttt{Facebook} & 63392 & 816886\\
\end{tabular}
\end{ruledtabular}
\end{table}

As we know, online social networks are divided into two types:
knowledge-sharing oriented and networking oriented~\cite{guolei-kdd09}.
For the data sets we use, \texttt{YouTube} belongs to the former, while
\texttt{Facebook} belongs to the latter, both of which are scale-free
networks.

\section{Structural role of weak ties}\label{sec:sroleofweakties}
\begin{figure*}[ht]
\centering \subfloat[\texttt{Facebook}]
{\label{fig:facebook_average_s}
\begin{minipage}[t]{0.5 \textwidth}
    \centering
    \includegraphics[scale=0.45]{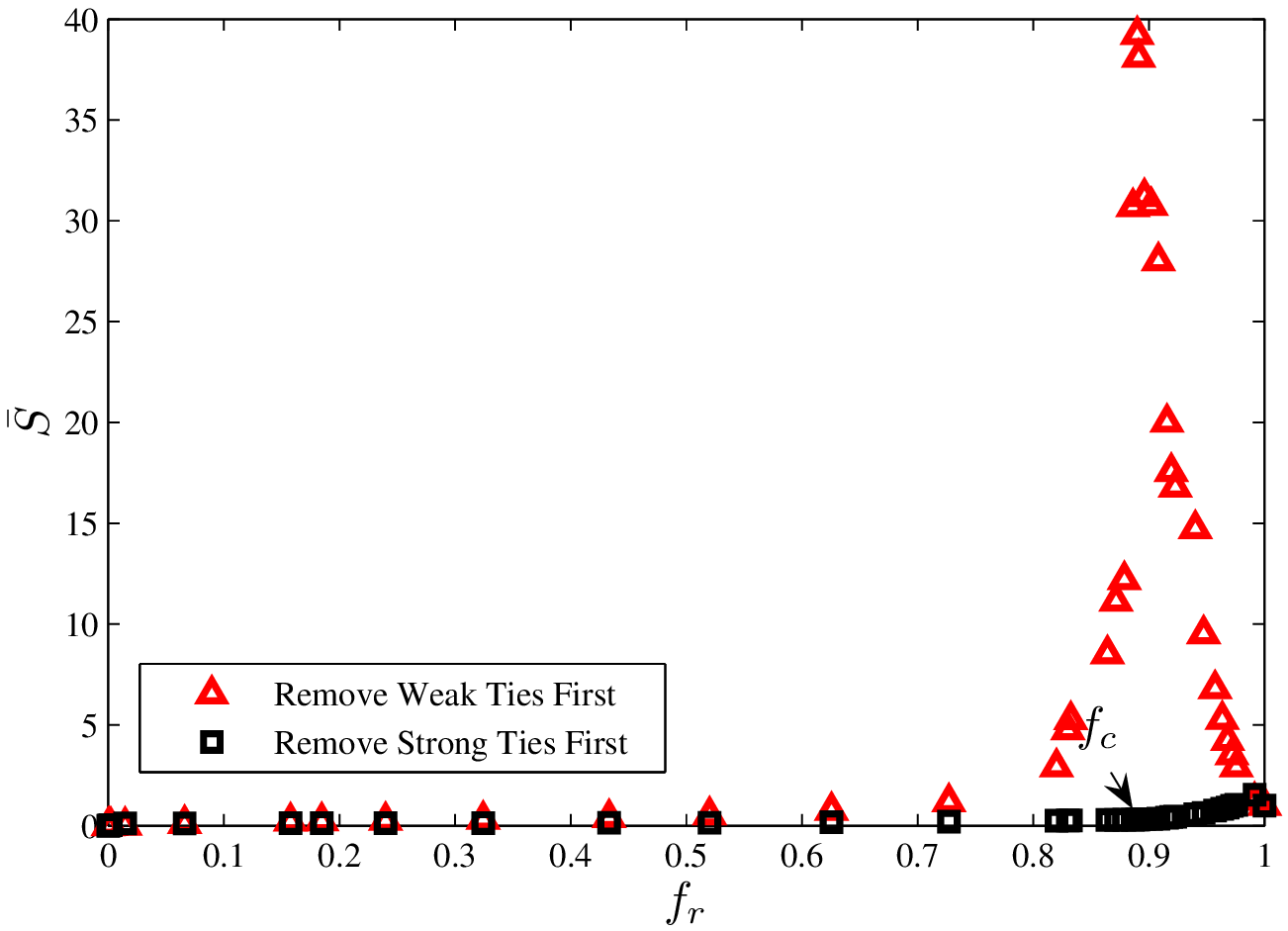}
\end{minipage}
} \subfloat[\texttt{Facebook}]{\label{fig:facebook_gcc}
\begin{minipage}[t]{0.45 \textwidth}    \centering
    \includegraphics[scale=0.45]{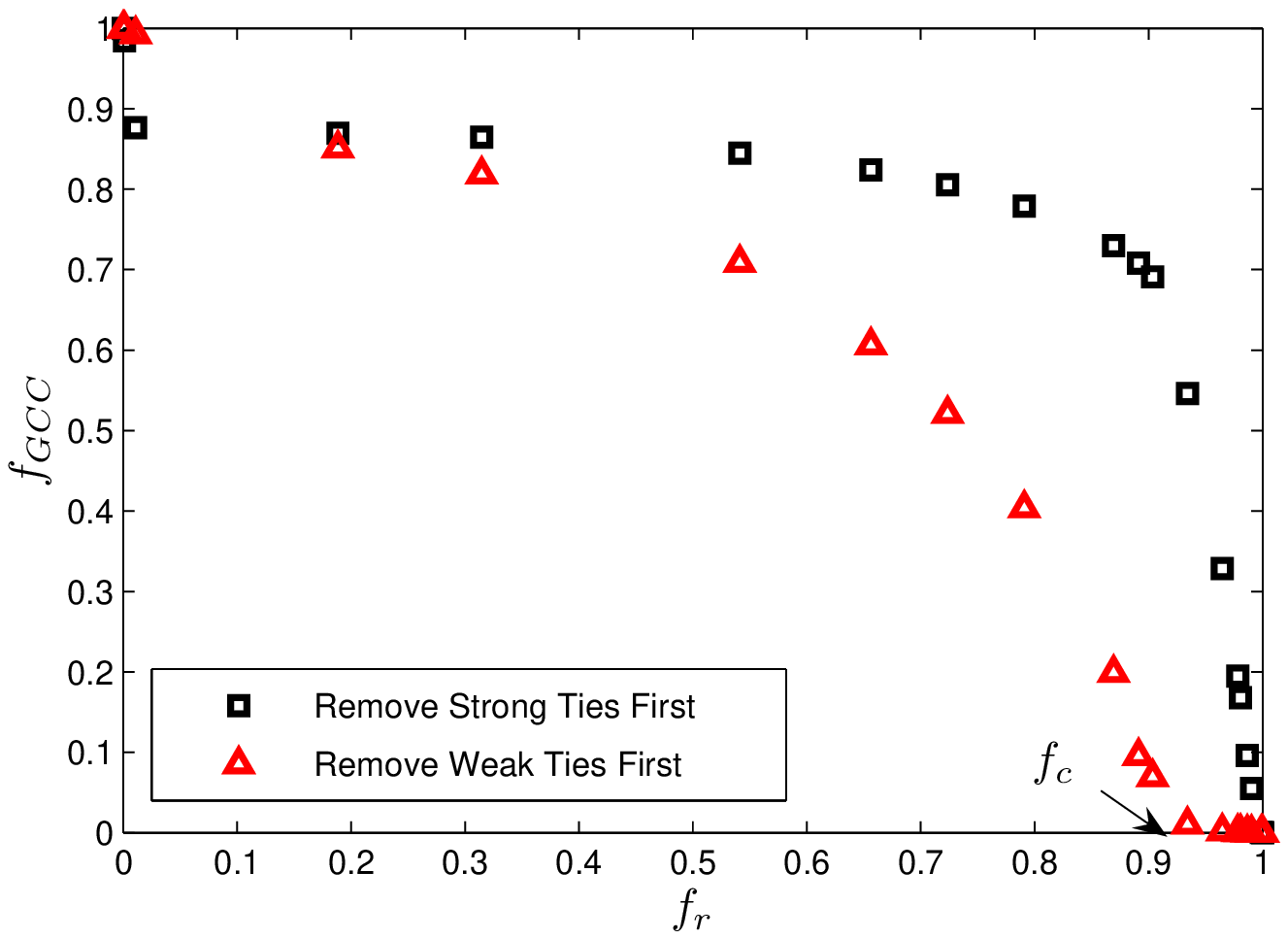}
\end{minipage}}\\
\subfloat[\texttt{YouTube}]{\label{fig:youtube_average_s}
\begin{minipage}[t]{0.45 \textwidth}
    \centering
    \includegraphics[scale=0.45]{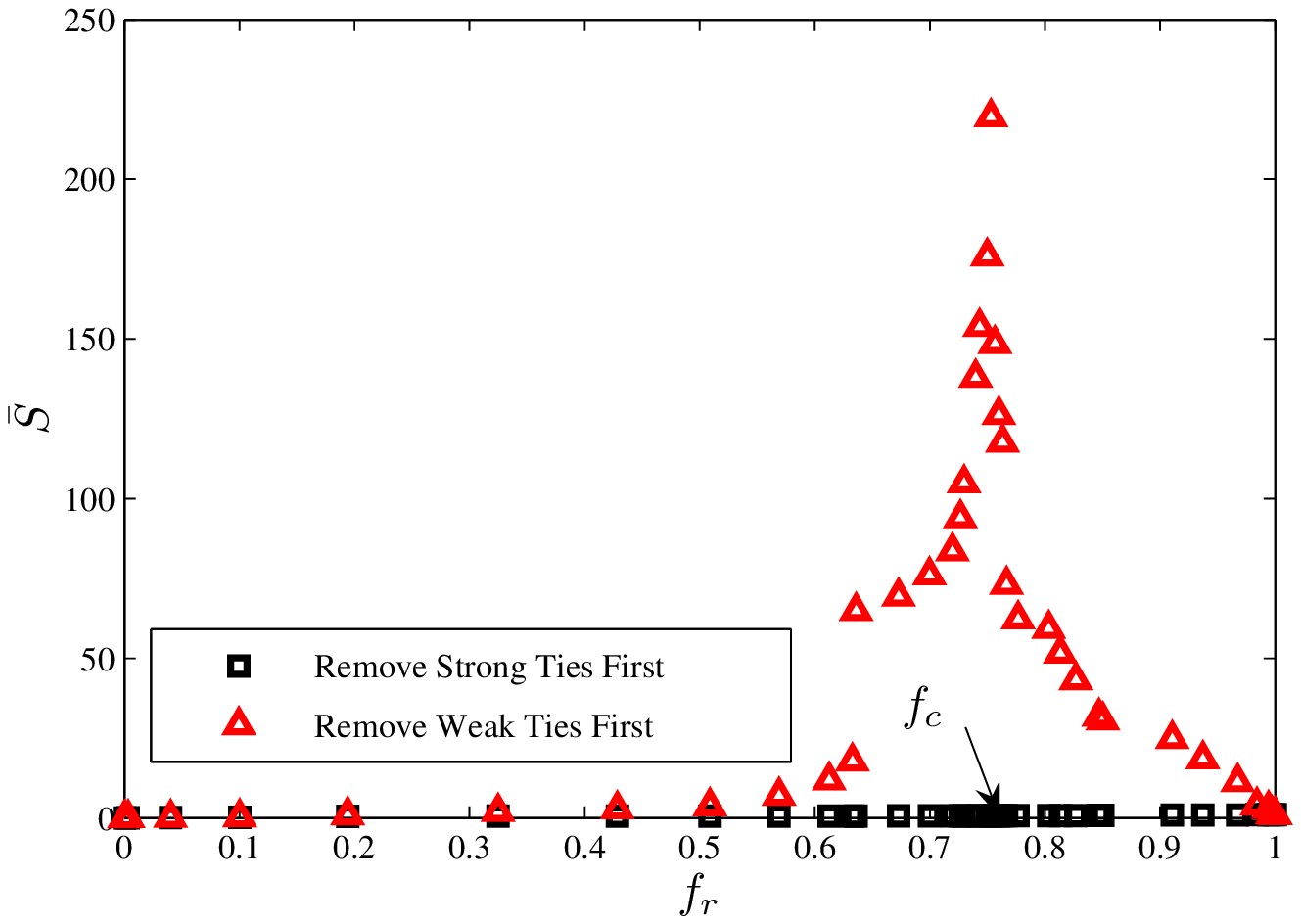}
\end{minipage}}
\subfloat[\texttt{YouTube}]{\label{fig:youtube_gcc}
\begin{minipage}[t]{0.45 \textwidth}
    \centering
    \includegraphics[scale=0.45]{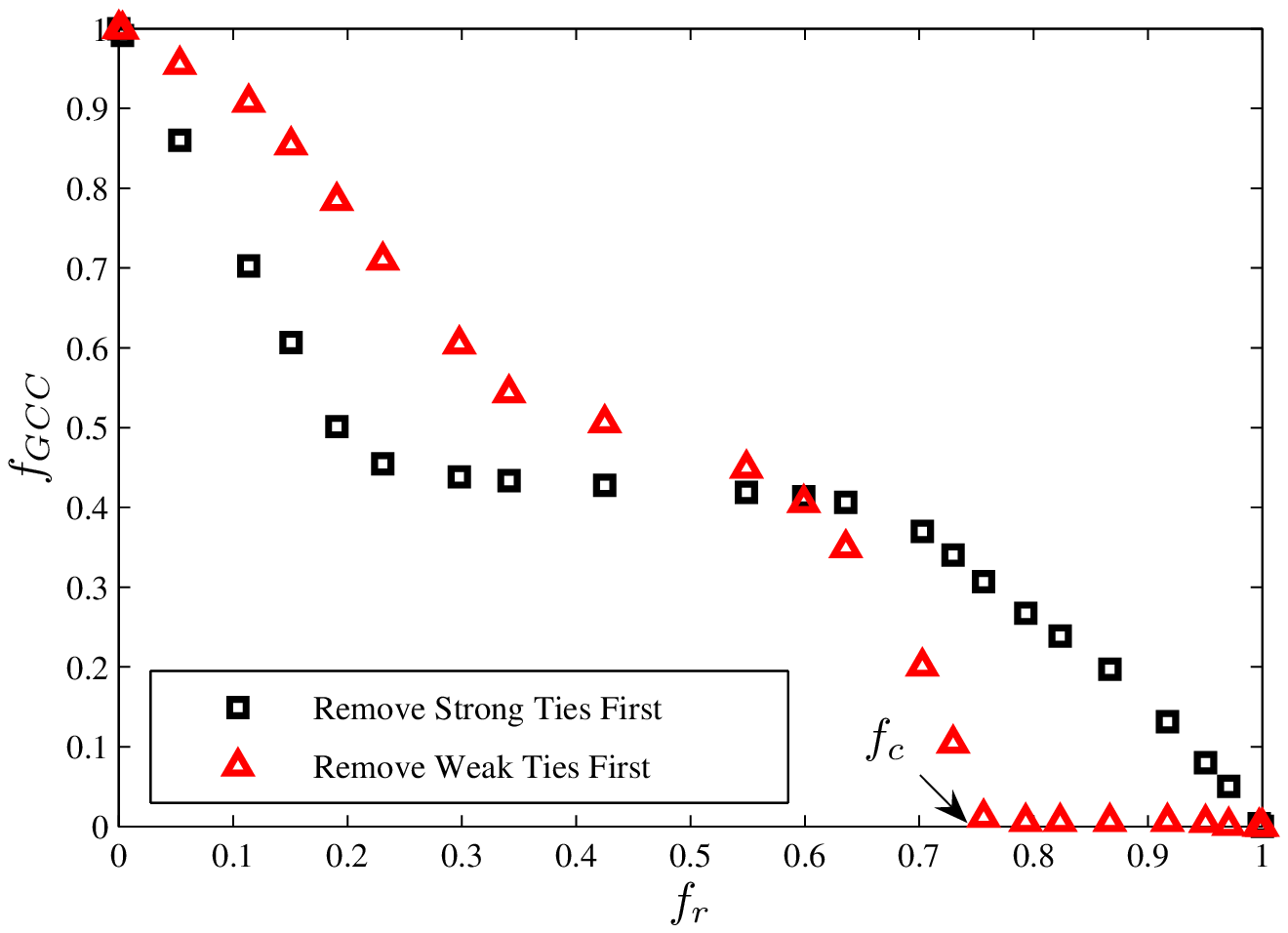}
\end{minipage}}
\caption{\label{fig:vofaverages}(Color online) The variations of
$\bar{S}$ and $f_{GCC}$ during the removal of weak ties first and
strong ties first, respectively. $f_r$ is the fraction of removed
ties.}
\end{figure*}

In this section, we study the structural role of weak ties. As shown
in Fig.~\ref{fig:facebook_average_s} and
Fig.~\ref{fig:youtube_average_s}, we find a phase transition
(characterized by $\bar{S}$) similar to the one
in~\cite{mobile-network} in online social networks during the
removal of weak ties first. This phase transition, however,
disappears if we remove the strong ties first. Furthermore, it is
also found in Fig.~\ref{fig:facebook_gcc} and
Fig.~\ref{fig:youtube_gcc} that the relative size of giant connected
cluster (GCC), denoted by $f_{GCC}$, shows different dynamics
between the removals of weak ties first and strong ties first.
We denote the critical fractions of the removed ties at the phase
transition point by $f_c$. It is interesting to note that
$f_c=0.753$ for \texttt{YouTube} and $f_c= 0.890$ for
\texttt{Facebook} when $\bar{S}$ reaches the submit, which are very
close to the case when $f_{GCC}\approx 0$.

In the percolation theory, the existence of the above phase transition
means that the network is collapsed, while the network is just shrinking
if there is no phase transition when removing the
ties~\cite{mobile-network}. So the above experiments tell us that weak
ties play a special role in the structure of online social networks, which
is different from the one strong ties play. In fact, they act as the
important bridges that connect isolated communities. In what follows, we
build a model that associates the weak ties with the information
diffusion, to discuss the coupled dynamics of the structure and the
information diffusion.

\section{\label{sec:droleofweakties}Diffusing role of weak ties}
The information diffusing in online social networks includes blogs,
photos, messages, comments, multimedia files, states, etc. Because of the
privacy control and other features of online social sites, the mechanism
of the information diffusion in online social networks is different from
traditional models, such as SIS, SIR and random walk. We start by
discussing the procedure of information diffusion in online social
networks.

\subsection{The Procedure of Information Diffusion}
The procedure of the diffusion in online social networks can be briefly
described as follows:
\begin{itemize}
\item
    The user $i$ publishes the information $I$, which may be a photo, a
blog, etc.
\item
    Friends of $i$ will know $I$ when they access the profile page of $i$ or
get some direct notifications from the online social site. We call
this scheme as \emph{push}.
\item
    Some friends of $i$, may be one, many or none, will comment, cite or
reprint $I$, because they think that it is interesting, funny or
important. We call this behavior as \emph{republish}.
\item
    The above steps will be repeated with $i$ replaced by each of those
who have republished $I$.
\end{itemize}

It is easy to find that the key feature of the information diffusion in
online social networks is that the information is pushed actively by the
site and only part of friends will republish it. Take \texttt{Facebook} as
an example, in which \emph{News Feed} and \emph{Live Feed} are two
significant and popular features. News Feed constantly updates a user's
profile page to list all his or her friends' news in \texttt{Facebook}.
The news includes conversations taking place between the walls of the
user's friends, changes of profile pages, events, and so
on~\cite{facebook-feature}. Live Feed facilitates the users to access the
details of the contents updated by News Feed. It is updated in a real-time
manner after the user's login to the web~\cite{facebook-help}. In fact,
News Feed aggregates the most interesting contents that a user's friends
are posting, while Live Feed shows to the user all the actions his or her
friends are taking in \texttt{Facebook}~\cite{news-feed}.

The feature of pushing and republishing we have discussed above is indeed
more obvious in \texttt{Twitter}, in which all the words you post will be
pushed immediately to your followers' terminals, including a PC or even a
mobile phone, and then they can republish it if they like. However, in
real-world situations, the trace of the information is hard to
collect~\cite{measurement-flick}, especially for large-scale networks. So
it is quite reasonable to build a model to characterize the mechanism and
simulate the diffusion.

\subsection{The Model for Information Diffusion}

Based on the procedure described above, we propose a simple model
$ID(\alpha,\beta)$, where $\alpha$ is the navigating factor and $\beta$
represents the strength of the information. In this model, $\alpha$
determines how to select neighbors to republish the information, while
$\beta\in[0,1]$ is a physical character of the information, which
describes how interesting, novel, important, funny or resounding it is.
The model is defined as follows:
\begin{itemize}
\item
    Step 1: Suppose there comes information $I$. Set the state of all the
nodes in $V$ to $\sigma_{0}$. The state $\sigma_{0}$ of a node means
$I$ is not known to it, otherwise the state is $\sigma_{1}$.
\item
    Step 2: Randomly select a seed node $i$ from the network. The degree
of $i$ is $k_i$. Set $i$ to $\sigma_{1}$. It publishes the information $I$
with strength equal to $\beta$ at time $T=0$.
\item
    Step 3: Increase the time by one unit, i.e., $T=T+1$. Set each node in
the neighborhood of $i$ to $\sigma_{1}$. Add $i$ to the set of nodes
that have published $I$, denoted by $P$. So $P=P\cup\{i\}$.
\item
    Step 4: Calculate the number of nodes that will republish $I$ in the
next round:
\begin{equation}
\label{eq:rounds} R_{i}=k_i\beta.
\end{equation}
\item
    Step 5: Select one node $j$ from the neighborhood of $i$ with the
probability~\cite{pij=0}
\begin{equation}
\label{eq:probability}p_{ij}=\frac{w_{ij}^\alpha}{\sum_{m=1}^{k_i}w_{im}^\alpha}.
\end{equation}
If $j$ is not in $P$, then add it to the set of nodes that will
republish $I$ in the next round, denoted by $W$. So $W=W\cup\{j\}$.
Repeat this step for $R_i$ times.
\item
    Step 6: For each node in $W$, execute from Step 3 to Step 5
recursively until $W$ is null or all the nodes in $V$ have known
$I$.
\end{itemize}

It is easy to find from Eq.~(\ref{eq:rounds}) that during the diffusion,
the number of republishing nodes selected from the neighborhood of $i$ is
decided by $k_i$ and $\beta$. It is consistent with the real situation
that the user with more friends tends to attract more other users to visit
and republish the information. The more interesting or important the
information is, the higher the chance that it will be republished. We use
parameter $\alpha$ in Eq.~(\ref{eq:probability}) to associate the
diffusion with the strength of the ties, which means different values of
$\alpha$ will lead to different selections of ties as paths for
republishing information in the next round. In fact, when $\alpha=-1$,
weak ties are to be selected preferentially as paths for republishing. The
selection is random when $\alpha=0$, and the strong ties will be selected
with higher priority when $\alpha=1$.

\subsection{Results and Analysis}

\begin{figure*}
\centering \subfloat[\texttt{Facebook}] {\label{fig:facebook_id_c}
\begin{minipage}[t]{0.45 \textwidth}
    \centering
    \includegraphics[scale=0.45]{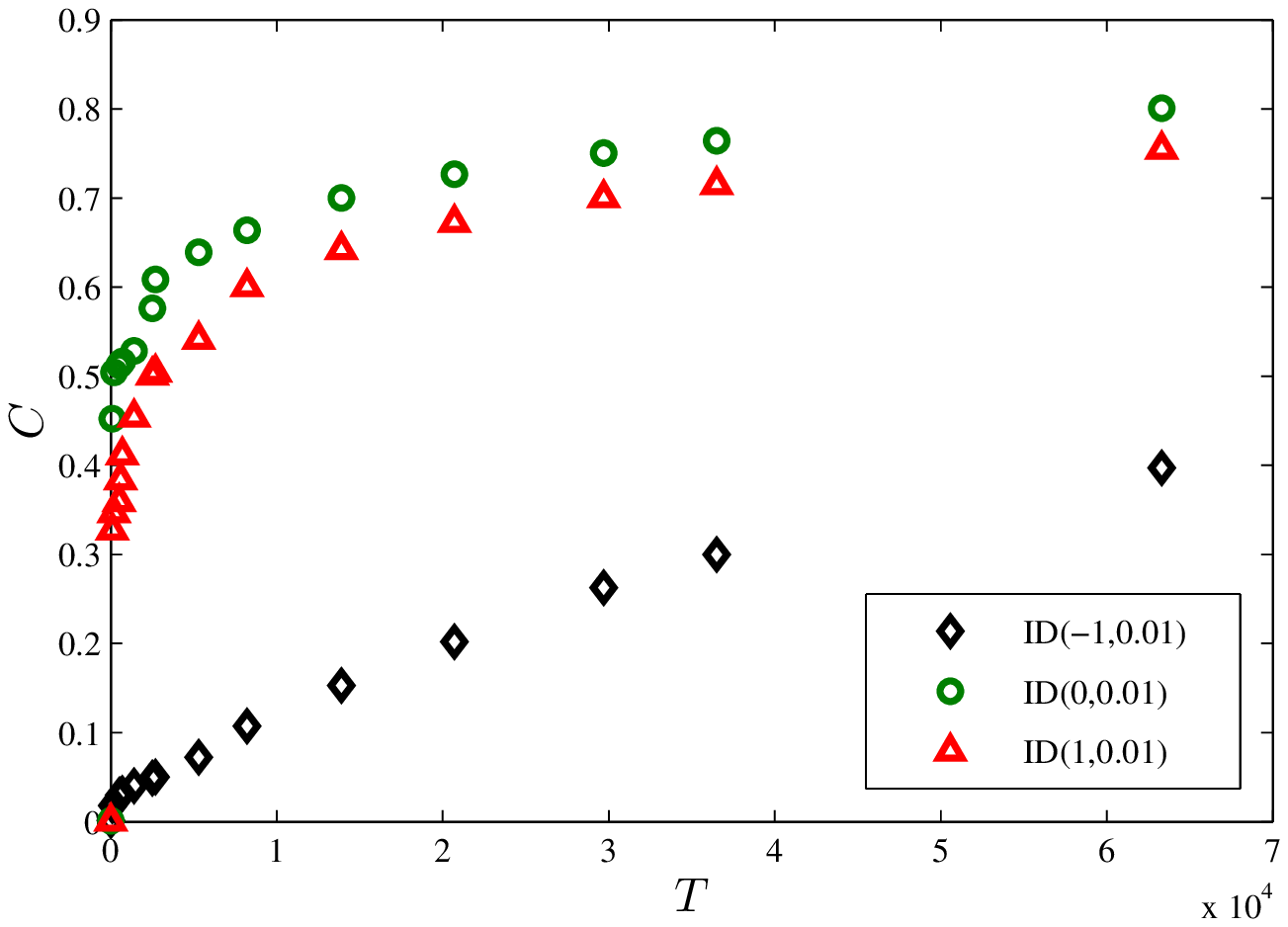}
\end{minipage}
} \subfloat[\texttt{YouTube}]{\label{fig:youtbe_id_c}
\begin{minipage}[t]{0.45 \textwidth}
    \centering
    \includegraphics[scale=0.45]{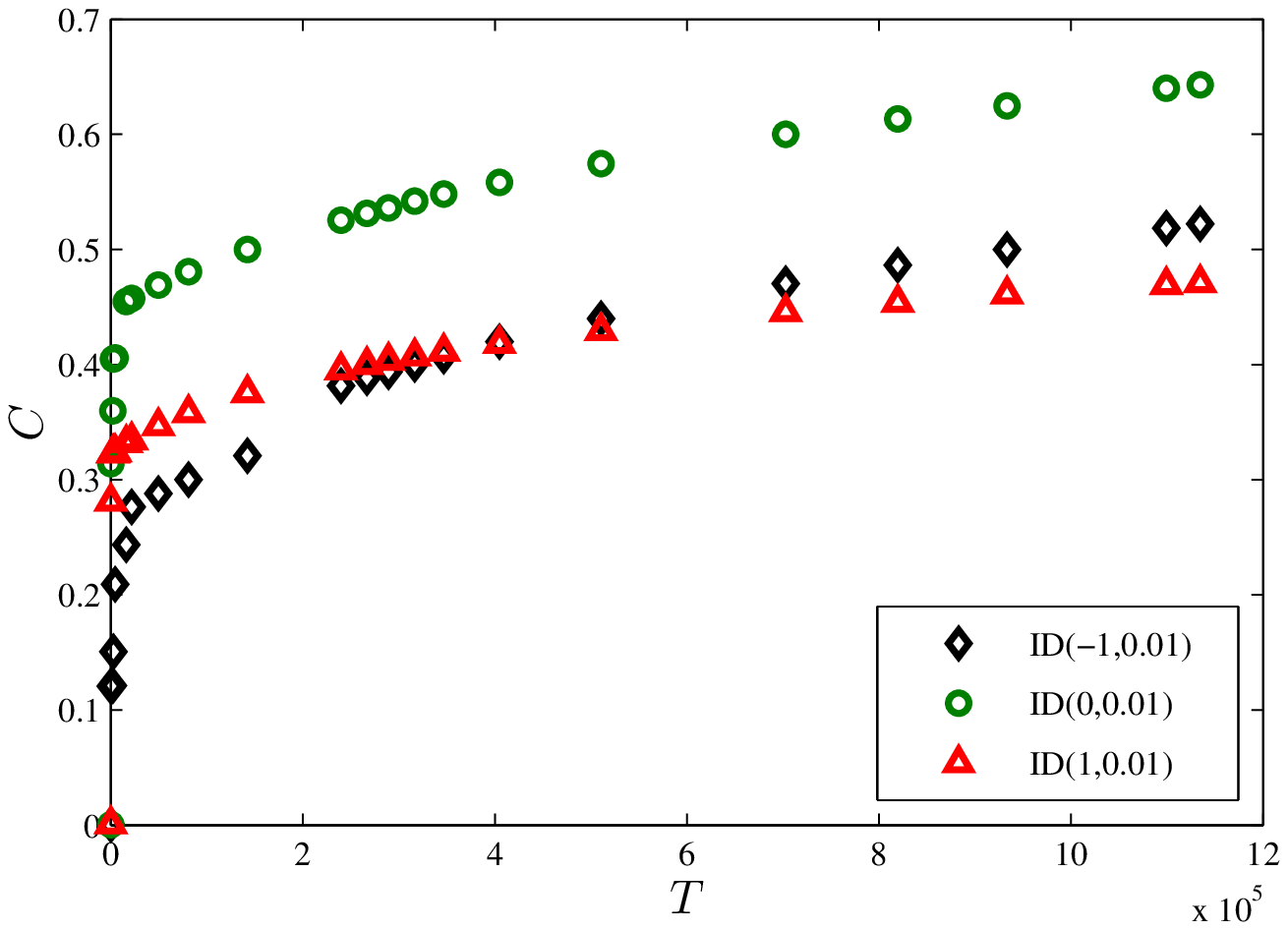}
\end{minipage}}
\caption{\label{fig:id_visited_t}(Color online) The dynamics of $C$ during
the process of the diffusion. We perform the experiments for each pair of
$\alpha$ and $\beta$ 20 times and return the mean value as the final
result.}
\end{figure*}

We define the fraction of nodes with the state $\sigma_{1}$ as the
coverage of $I$, denoted by $C$. Since it is found that only 1-2\%
friends will republish the information in
Flickr~\cite{measurement-flick}, we let $\beta=0.01$ in the
simulations. Fig.~\ref{fig:id_visited_t} shows the numeric
experimental results on \texttt{Facebook} and \texttt{YouTube}
networks. As can be seen, $C$ reaches the maximum when $\alpha=0$.
In other words, compared with weak or strong ties, selecting the
republishing nodes randomly from the neighborhood will make the
information spread faster and wider. This is indeed out of our
expectation, since previous studies show that weak ties can
facilitate the information diffusion in social networks.

To understand this, we further explore the process of the information
diffusion in details. By Eq.~(\ref{eq:tiestrength}), we can easily have
$$1/w_{ij}=(k_i-2)/c_{ij}+k_j/c_{ij}-1.$$ Assume that as $k_j$
increases, $c_{ij}$ increases proportionately, i.e., $k_j/c_{ij}=const$.
Then given a node $i$ and its neighbor node $j$, we have
$k_j\uparrow\Rightarrow c_{ij}\uparrow\Rightarrow
1/w_{ij}\downarrow\Rightarrow w_{ij}\uparrow$, and vice versa. This
implies that a neighbor node of $i$ tends to have a higher degree if it
has a stronger strength of ties with $i$. Therefore, when selecting the
republishing nodes for the next round from the neighborhood, different
$\alpha$ will select nodes with different degrees preferentially. For
example, when $\alpha=-1$, the weak ties will be selected with higher
priority, which means that the nodes with lower degrees will be selected
preferentially. However, it is easy to learn from Eq.~(\ref{eq:rounds})
that, for the node with lower degree, the republishing nodes selected from
its neighborhood will be less, which will eventually reduce the total
number of republishing nodes and impede the information from further
spreading in the network. As to the case of selecting strong ties
preferentially, although it will tend to select the nodes with higher
degrees to republish, the local trapping~\cite{mobile-network} will limit
the scope of selected nodes into some local areas and make it harder to
propagate the information further in the network.

\begin{figure*}
\centering \subfloat[\texttt{Facebook}] {\label{fig:facebook_fpub}
\begin{minipage}[t]{0.45 \textwidth}
    \centering
    \includegraphics[scale=0.45]{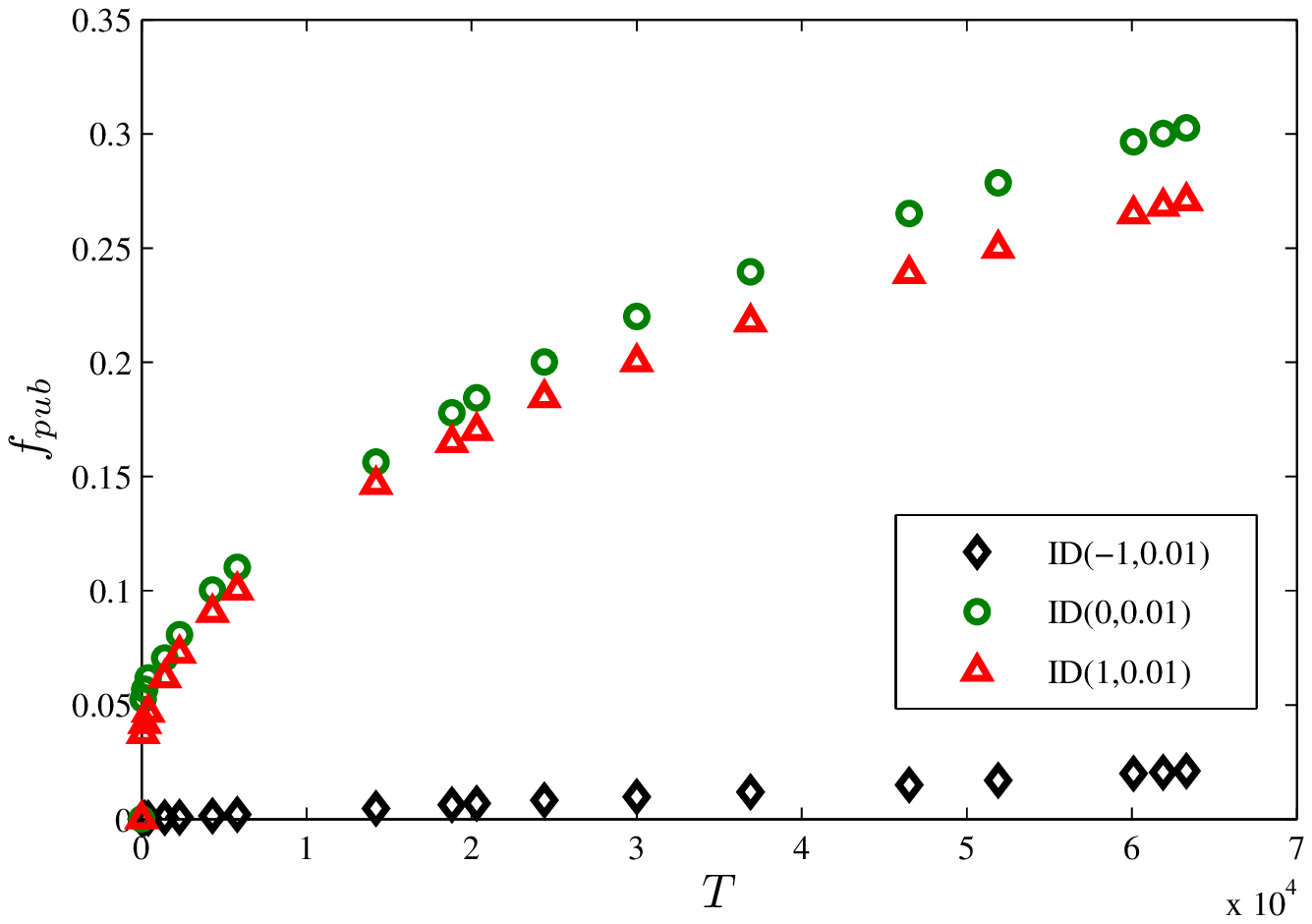}
\end{minipage}
} \subfloat[\texttt{YouTube}]{\label{fig:youtbe_fpub}
\begin{minipage}[t]{0.45 \textwidth}
    \centering
    \includegraphics[scale=0.45]{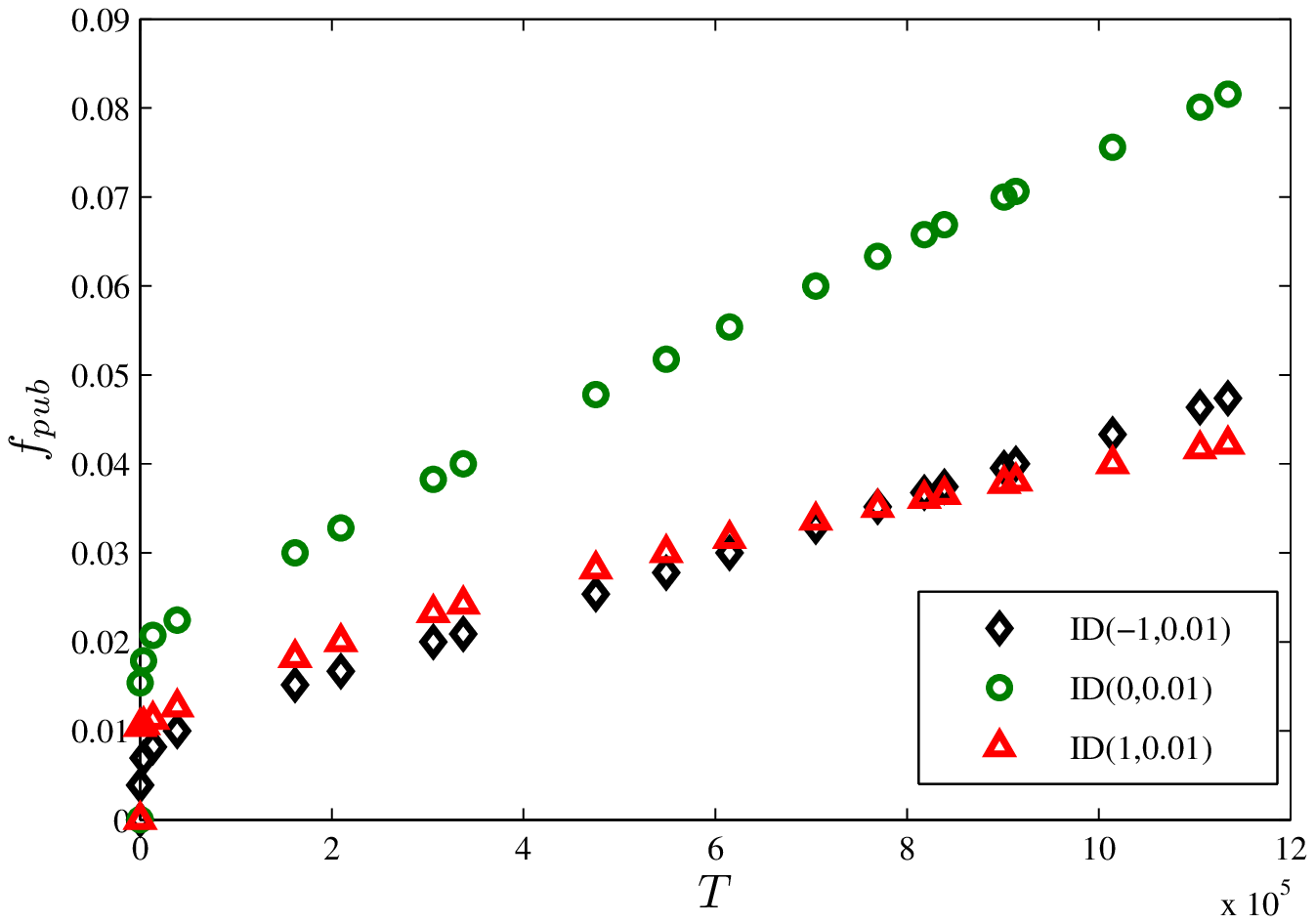}
\end{minipage}}
\caption{\label{fig:fpub}(Color online) The dynamics of $f_{pub}$ during
the process of the diffusion. We perform the experiments for each pair of
$\alpha$ and $\beta$ 20 times and return the mean value as the final
result.}
\end{figure*}

\begin{figure*}
\centering \subfloat[\texttt{Facebook}] {\label{fig:facebook_flocal}
\begin{minipage}[t]{0.45 \textwidth}
    \centering
    \includegraphics[scale=0.45]{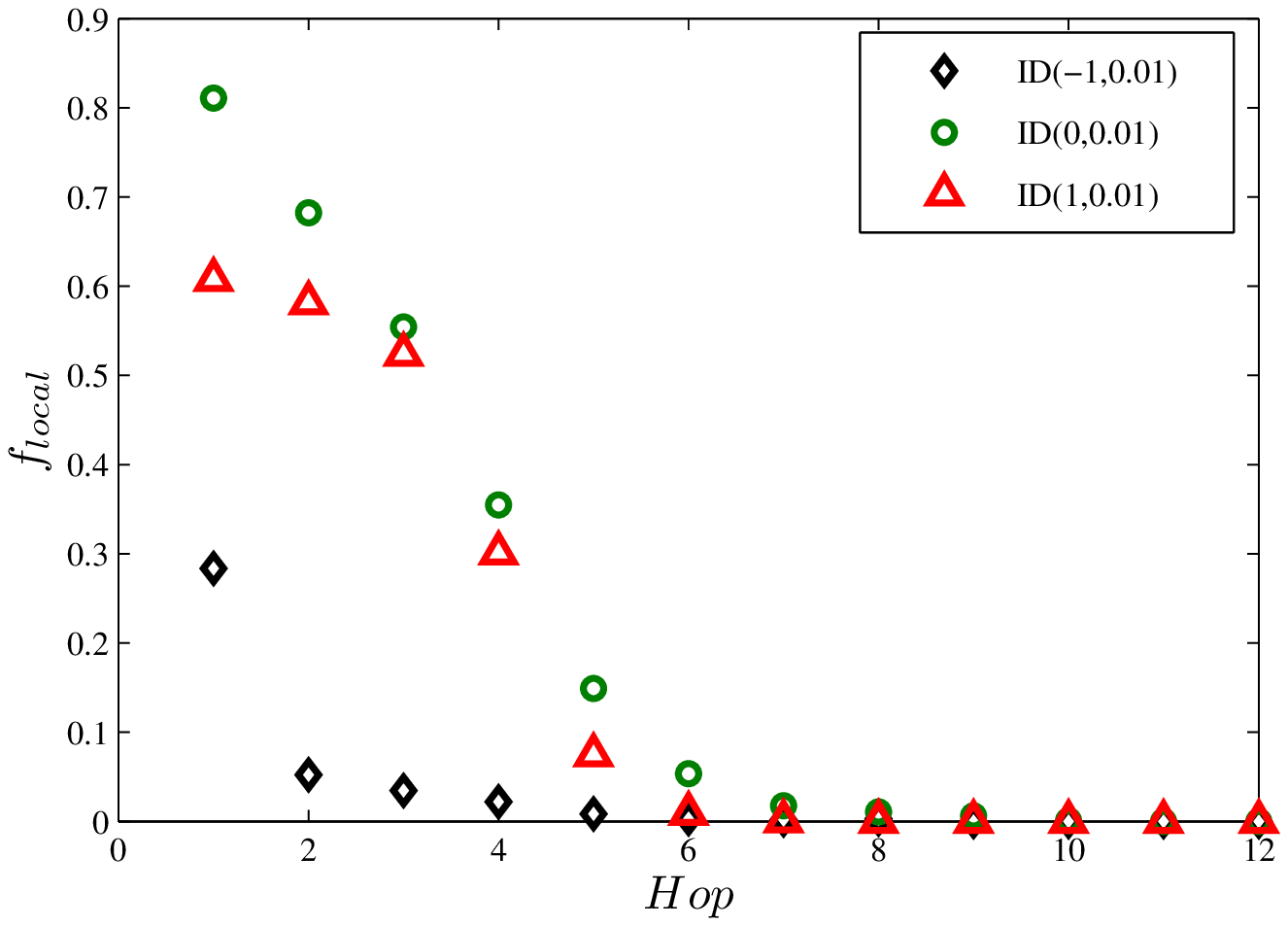}
\end{minipage}
} \subfloat[\texttt{YouTube}]{\label{fig:youtbe_flocal}
\begin{minipage}[t]{0.45 \textwidth}
    \centering
    \includegraphics[scale=0.45]{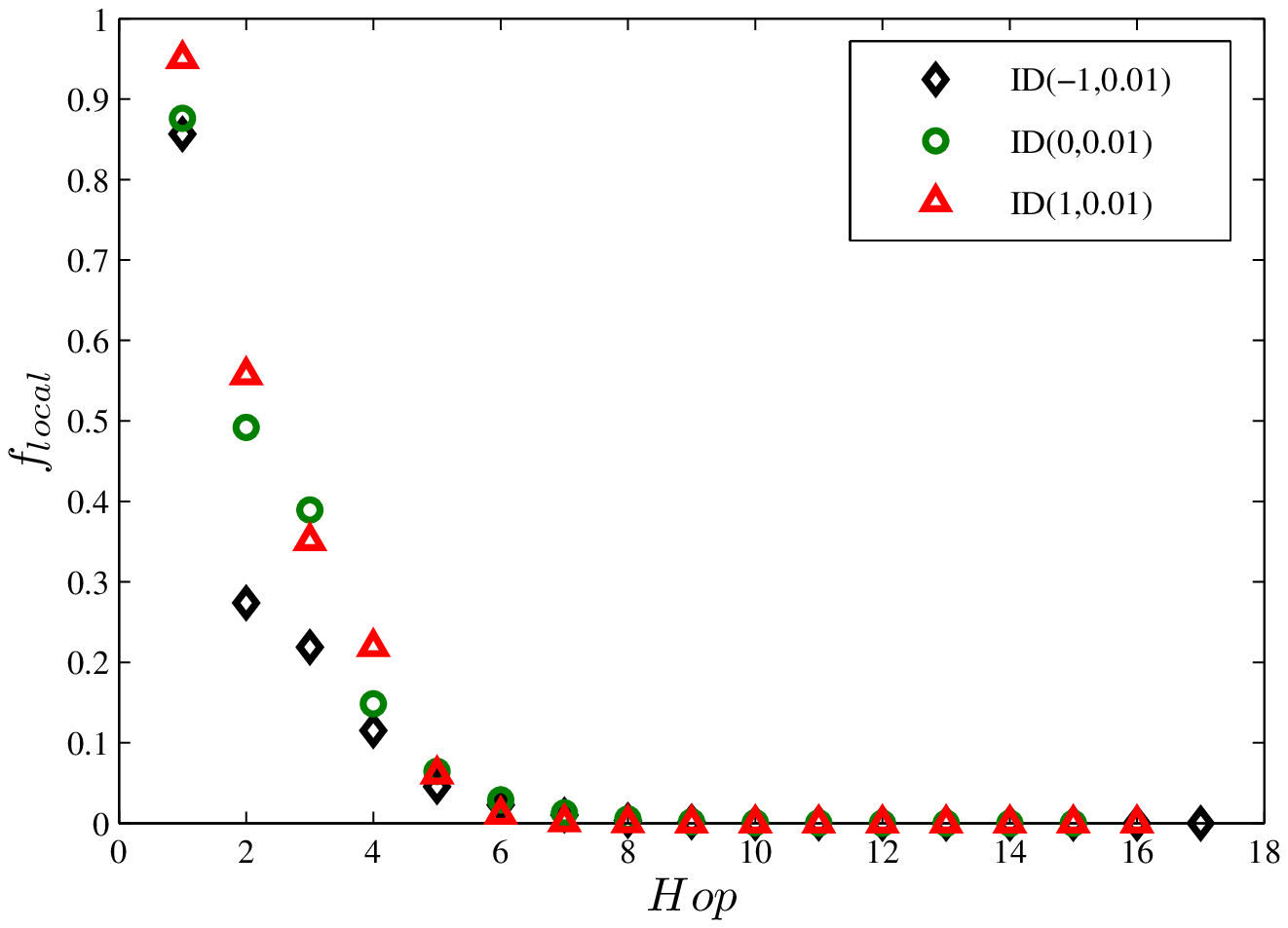}
\end{minipage}}
\caption{\label{fig:flocal}(Color online) The dynamics of $f_{local}$
during the information propagation far away from the source. We perform
each experiment 20 times and get the mean value as the final result.}
\end{figure*}

To validate the analysis above, we also observe the fraction of the nodes
that have published $I$ during the diffusion, denoted by $f_{pub}$. As
shown in Fig.~\ref{fig:fpub}, $f_{pub}$ increases more slowly when
$\alpha=-1$, and the time-varying properties of $f_{pub}$ are similar to
those of $C$ in Fig.~\ref{fig:id_visited_t} for different $\alpha$ values,
respectively. We also monitor the fraction of the nodes that have
published $I$ in each hop away from the source node, denoted by
$f_{local}$. As shown in Fig.~\ref{fig:flocal}, when $\alpha=-1$,
$f_{local}$ decreases faster than other cases, in particular the
$\alpha=0$ case. It means when $\alpha=-1$, the number of republishing
nodes selected from the neighborhood decreases sharply as the information
spreading far away from the source, which agrees with our former analysis.
As for the case of $\alpha=1$, $f_{pub}$ increases more and more slowly
during the diffusion, because the nodes selected to republish are trapped
in some local clusters. In other words, it is hard to find some new nodes
to republish the information to the outer space.

\begin{figure*}
\centering \subfloat[\texttt{Facebook}]
{\label{fig:facebook_remove_ties}
\begin{minipage}[t]{0.45 \textwidth}
    \centering
    \includegraphics[scale=0.45]{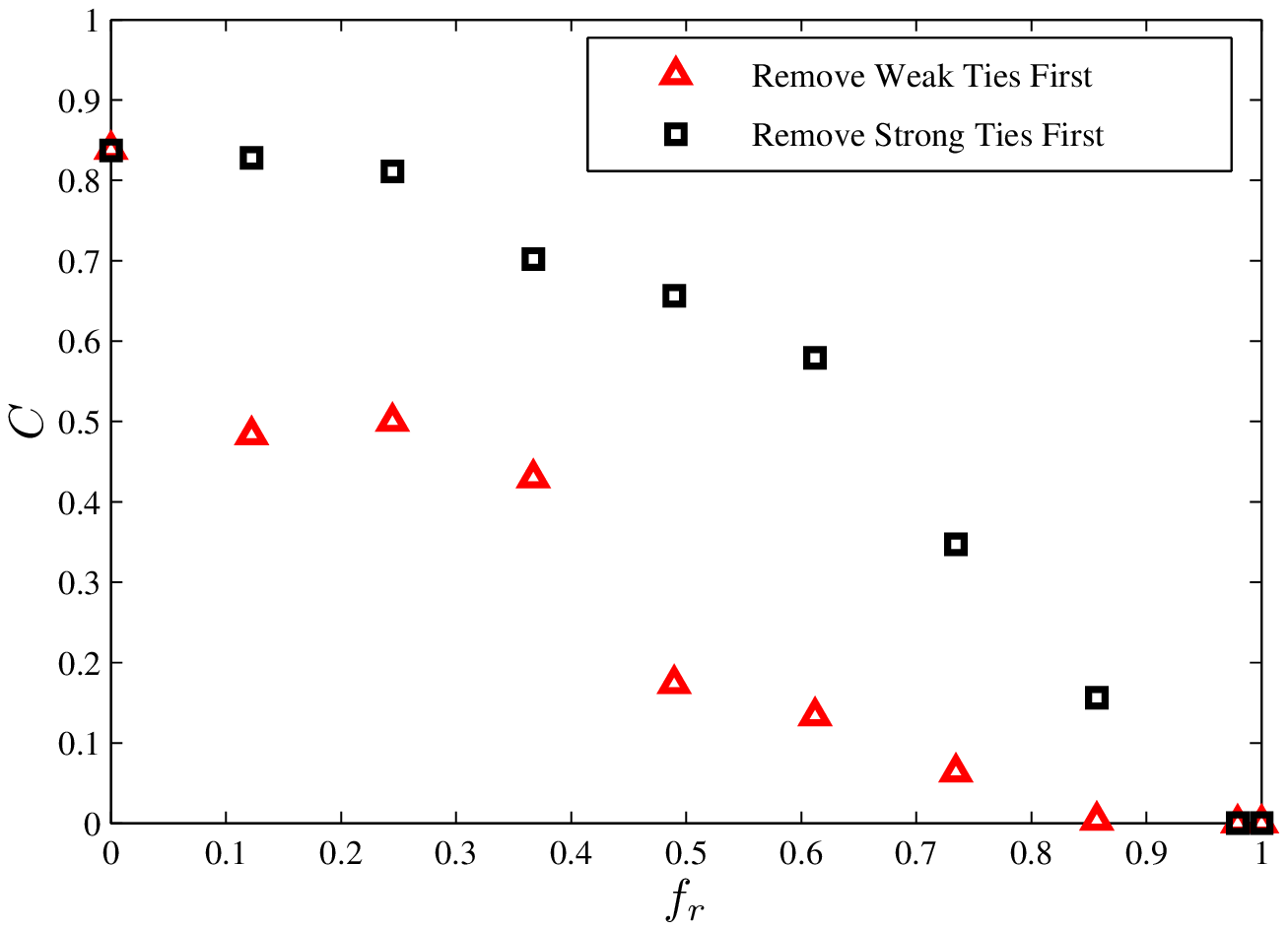}
\end{minipage}
} \subfloat[\texttt{YouTube}]{\label{fig:youtbe_remove_ties}
\begin{minipage}[t]{0.45 \textwidth}
    \centering
    \includegraphics[scale=0.45]{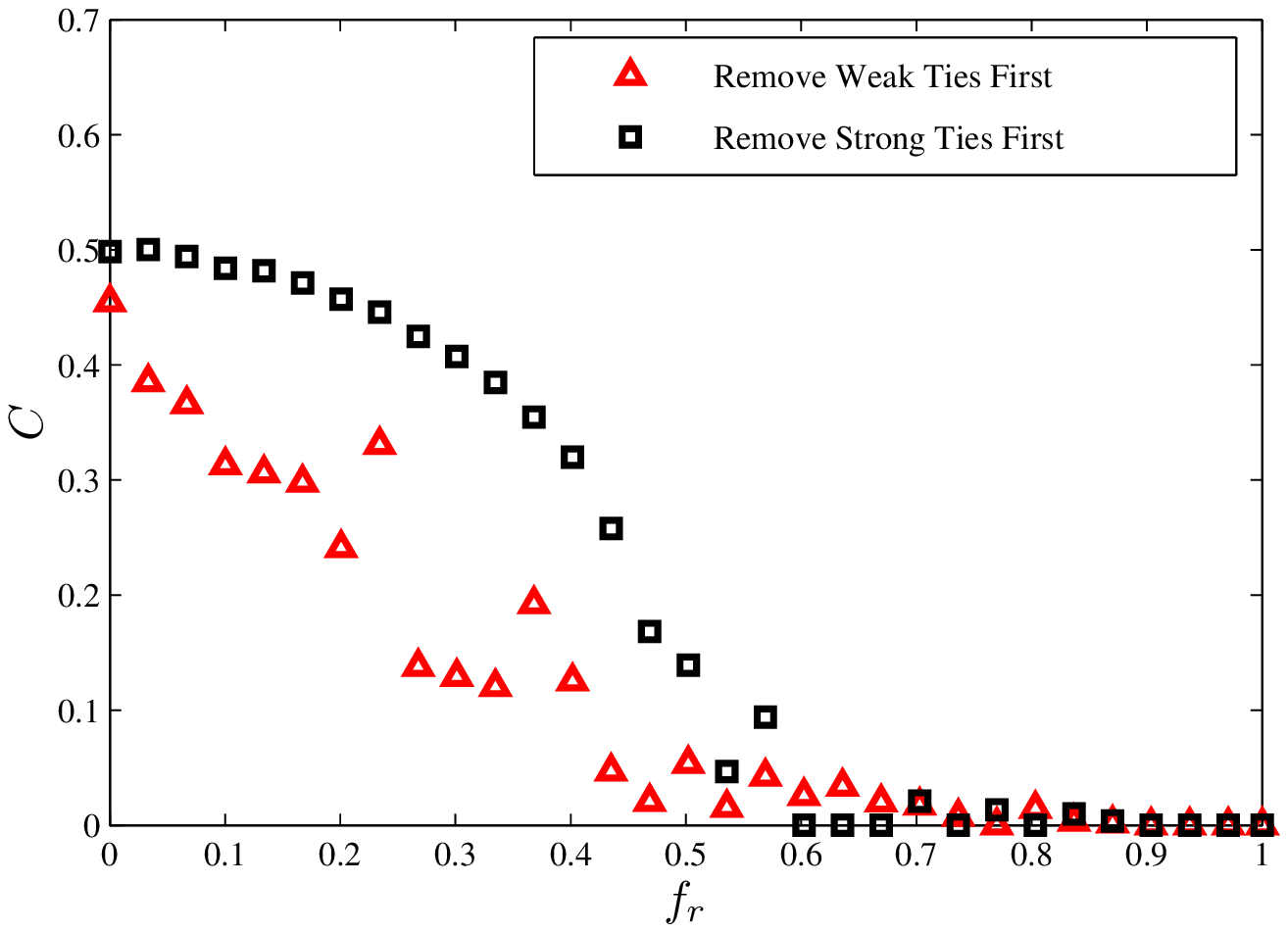}
\end{minipage}}
\caption{\label{fig:remove_ties}(Color online) The variations of $C$
during the removal of ties. The diffusing time is $T_{Facebook}=\arrowvert
V \arrowvert$ and $T_{YouTube}=10^4$. We perform the experiments 20 times
for $\alpha=0$ and $\beta=0.01$, and return the mean value as the final
result.}
\end{figure*}

Based on the above results, we can conclude that selecting weak ties
preferentially as the path to republish information cannot make it diffuse
faster. However, this does not mean that weak ties play a trivial role in
the information diffusion in online social networks, especially when we
recall its special role in the network structure in
Section~\ref{sec:sroleofweakties}. Let $\alpha=0$ in $ID(\alpha,\beta)$,
we compare the variation of $C$ under the situation of removing weak ties
first with that of removing strong ties first. As shown in
Fig.~\ref{fig:remove_ties}, for the case of removing weak ties first, the
coverage of the information decreases rapidly, e.g., from 0.8 to 0.4 in
\texttt{Facebook} when the fraction of removed weak ties reaches about
0.4. This implies that weak ties are indeed crucial for the coverage of
information diffusion in online social networks.

\begin{figure*}
\centering \subfloat[\texttt{Facebook}] {\label{fig:facebook_beta}
\begin{minipage}[t]{0.45 \textwidth}
    \centering
    \includegraphics[scale=0.45]{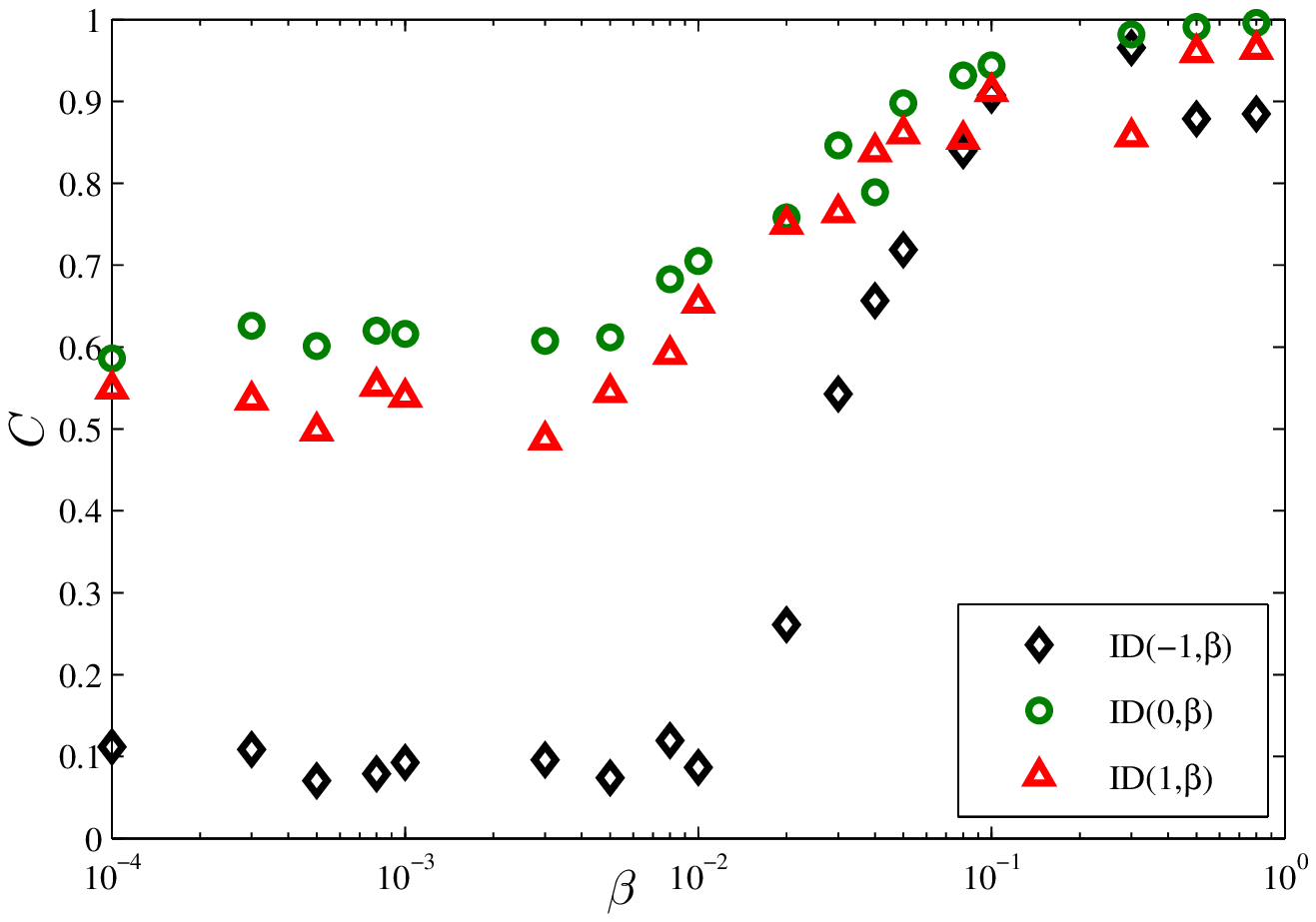}
\end{minipage}
} \subfloat[\texttt{YouTube}]{\label{fig:youtbe_beta}
\begin{minipage}[t]{0.45 \textwidth}
    \centering
    \includegraphics[scale=0.45]{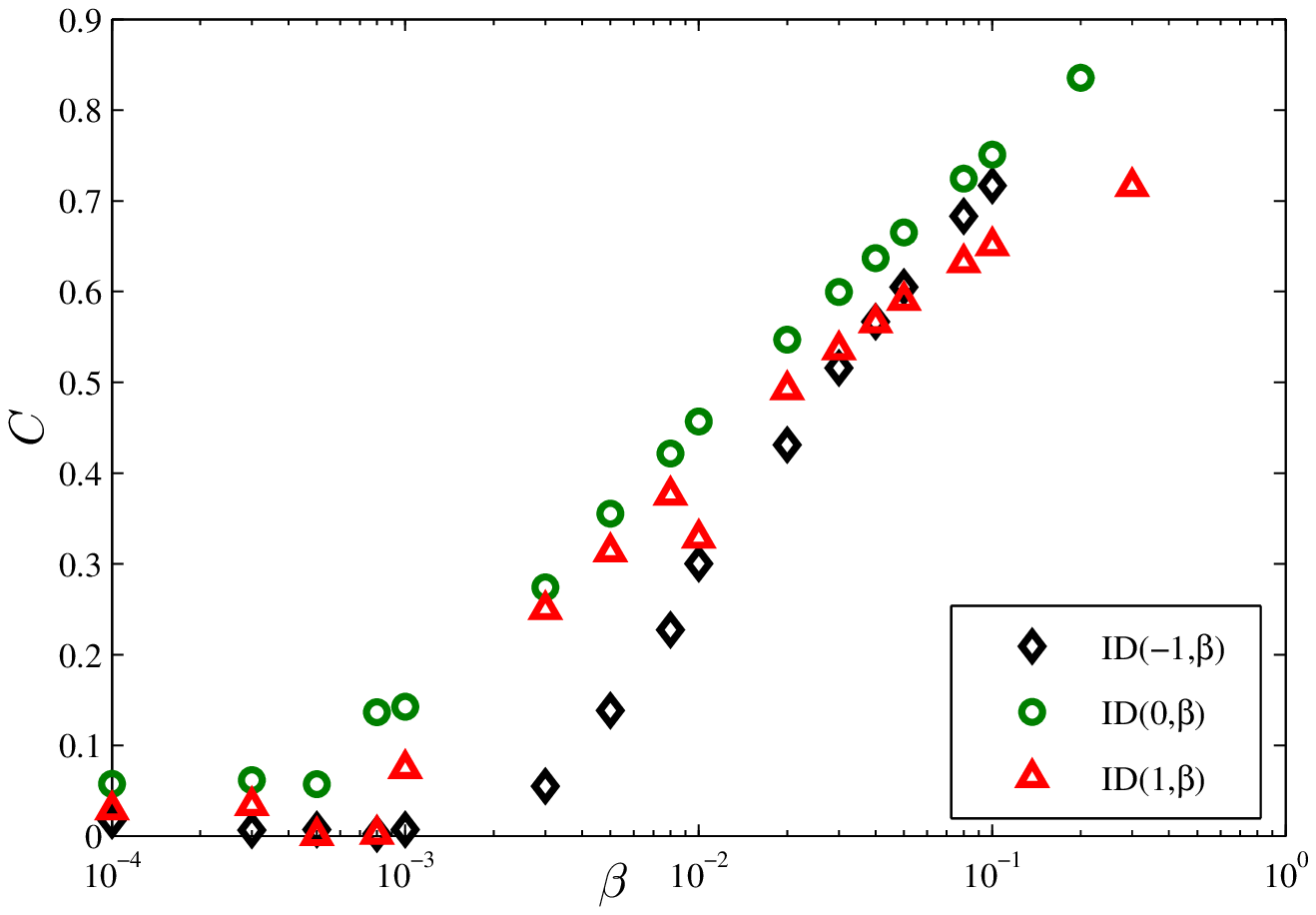}
\end{minipage}}
\caption{\label{fig:coverage_beta}(Color online) The increment of $C$ when
$\beta$ grows in the log-scale. We perform the experiments for each pair
of $\alpha$ and $\beta$ 20 times and return the mean value as the final
result.}
\end{figure*}

To further study the effect of $\beta$, we conduct experiments with
different $\beta$ values, as shown in Fig.~\ref{fig:coverage_beta}. As can
be seen, no matter what the $\beta$ value is, random selection
($\alpha=0$) is still the fastest mode for the information diffusion,
although the gap tends to shrink with higher $\beta$ values. It is also
shown that when $\beta$ grows, $C$ will also rise for all $\alpha$ values.
That is, the greater the strength of the information is, the more nodes
will be attracted to republish it, and the wider it will spread in the
network.

Until now we can conclude that weak ties play a subtle role in the
information diffusion in online social networks. On one hand, they are
bridges that connect isolated communities and break through the trapping
of information in local areas~\cite{mobile-network}. On the other hand,
selecting weak ties preferentially as the path of republishing cannot make
the information diffuse faster and wider.

\section{\label{sec:dcontrol}Diffusion Control}

The growing popularity of the online social networks does not mean
that it is safe and reliable. On the contrary, the virus spread and
the private information diffusion have made it become a massive
headache for IT administrators and users~\cite{virus-attack,
security-risk}. For example, ``KooFace'' is a Trojan Worm on
Facebook, which spreads by leaving a comment on profile pages of the
victim's friends to trap a click on the malicious
link~\cite{facebook-spread}. About 63\% of system administrators
worry that their employees will share too much private information
online~\cite{business-fear}. So as time goes by, it becomes more and
more important and urgent to control the virus spread and the
private information diffusion in online social networks.

In the light of this, we can make use of the weak ties for the information
diffusion control. That is, in the real-world practices, we can assume
that the behavior of republishing information is random, i.e., $\alpha=0$.
Then according to the results in Fig.~\ref{fig:remove_ties}, we can make
the virus or the private information trapped in local communities by
removing weak ties and stop them from diffusing further in the network.

\section{\label{sec:summary}Summary}

Online social sites have become one of the most popular Web 2.0
applications in the Internet. As a new social media, the core
feature of online social networks is the information diffusion. We
investigate the coupled dynamics of the structure and the
information diffusion in the view of weak ties. Different from the
recent work~\cite{measurement-flick}, we do not focus on the trace
collection and analysis of the real data flowing in the network.
Instead, inspired by~\cite{mobile-network}, we propose a model for
online social networks and take a closer look at the role of weak
ties in the diffusion.

We find that the phase transition found in the mobile communication
network exists pervasively in online social networks, which means
that the weak ties play a special role in the network structure.
Then we propose a new model $ID(\alpha,\beta)$, which associates the
strength of ties with the diffusion, to simulate how the information
spreads in online social networks. Contrary to our expectation,
selecting weak ties preferentially to republish cannot facilitate
the information diffusion in the network, while the random selection
can. Through extra analysis and experiments, we find that when
$\alpha=-1$, the nodes with lower degrees are preferentially
selected for republishing, which will limit the scope of the
distribution of republishing nodes in the following rounds. However,
even for the random selection case, removal of the weak tie can make
the coverage of the information decreases sharply, which is
consistent with its special role in the structure.

So we conclude that weak ties play a subtle role in the information
diffusion in online social networks. On one hand, they play a role of
bridges, which connect isolated communities and break through the trapping
of information in local areas. On the other hand, selecting weak ties
preferentially to republish cannot make the information diffuse faster in
the network. For potential applications, we think that the weak ties might
be of use in the control of the virus spread and the private information
diffusion.

\begin{acknowledgments}
This work was supported by National 973 Program of China (Grant
No.2005CB321901) and the fund of the State Key Laboratory of Software
Development Environment (SKLSDE-2008ZX-03). The second author was
supported partially by National Natural Science Foundation of China (Grant
No. 70901002 and 90924020) and Beihang Innovation Platform Funding (Grant.
No YMF-10-04-024).
\end{acknowledgments}


\begin{thebibliography}{34}
\bibitem[{pij()}]{pij=0}
\bibinfo{note}{In order to make {$p_{ij}>0$} for the case of {$w_{ij}=0$}, we
  set {$w_{ij}=1/2N$} (the smallest possible value of {$w_{ij}$} except zero),
  where {$N$} is the size of the network.}
\end{thebibliography}

\end{document}